\documentclass[10pt,prb,twocolumn,showpacs,amssymb,floatfix]{revtex4}
\usepackage{amsmath,graphics,epsfig,mathrsfs}
\usepackage{amsbsy}
\usepackage{bm}
\usepackage{epigraph}
\newcommand{\mac}{\mathcal}
\newcommand{\msc}{\mathscr}

\newcommand{\ti}{\textit}

\newcommand{\nn}{\nonumber}
\newcommand{\pat}{\partial}
\newcommand{\para}{\parallel}
\newcommand{\pr}{\prime}

\newcommand{\raw}{\rightarrow}

\newcommand{\ub}{\underbrace}

\newcommand{\alp}{\alpha}
\newcommand{\dlt}{\delta}
\newcommand{\Dlt}{\Delta}

\newcommand{\eps}{\epsilon}
\newcommand{\gm}{\gamma}

\newcommand{\og}{\omega}
\newcommand{\Og}{\Omega}
\newcommand{\sg}{\sigma}
\newcommand{\Sg}{\Sigma}

\newcommand{\etal}{{\em et al.,~}}
\newcommand{\ie}{{i.e.,~}}

\begin{document}
\title{Rashba spin torque in an ultrathin ferromagnetic metal layer}
\author{Xuhui Wang}
\email{xuhui.wang@kaust.edu.sa}
\author{Aurelien Manchon}
\email{aurelien.manchon@kaust.edu.sa}
\affiliation{Physical Science \& Engineering Division,
KAUST, Thuwal 23955-6900, Kingdom of Saudi Arabia}
\date{\today}

\begin{abstract}
In a two-dimensional ferromagnetic metal layer lacking inversion symmetry,
the itinerant electrons mediate the interaction between the
Rashba spin-orbit interaction and the ferromagnetic
order parameter, leading to a Rashba spin torque exerted on the magnetization.
Using Keldysh technique, in the presence of both magnetism and a spin-orbit
coupling, we derive a spin diffusion equation that provides a coherent
description to the diffusive spin dynamics. The characteristics of the spin
torque and its implication on magnetization dynamics are discussed in the limits of large and weak
spin-orbit coupling.
\end{abstract}
\pacs{75.60.Jk,75.70.Tj,72.25.-b,72.10.-d}
\maketitle

\section{Introduction}
By transferring angular
momentum between the electronic spin and the orbital, spin-orbit coupling
fills the need for electrical manipulation of spin degree of freedom.
Outstanding examples are the electrically generated bulk spin polarization \cite{dp-she-1971,edelstein-1989} and
the well-known spin Hall effect (SHE) \cite{murakami-science-2004,sinova-prl-2004,kato-science-2004}
in a two dimensional electron gas where the spin-orbit interaction,
particularly of the Rashba-type, \cite{rashba-soi} plays the leading role.
Rashba spin-orbit interaction not only introduces an effective field perpendicular to the
linear momentum but also provides the backbone
to the spin-relaxation through the so-called D'yakonov-Perel mechanism, \cite{dp} which
is dominant in a two-dimensional system.
Besides its prominent role in semiconductors, Rashba spin-orbit coupling
is believed to exist at ferromagnetic/heavy metal as well as
ferromagnetic/metal-oxide interfaces, in which the inversion symmetry
breaking offers a potential gradient empowering the spin-orbit coupling.

Meanwhile, magnetism continuously stimulates the industrial and academic appetite.
In the pursuit of fast magnetization switching,
Slonczewski-Berger spin transfer torque \cite{slonczewski-berger-1996}
employs a polarized spin current instead of a cumbersome magnetic field.
This celebrated scheme demands non-collinear
magnetic textures in forms of, for example, spin valves or domain wall structures.\cite{refstt}

In the presence of inversion symmetry breaking (such as asymmetric interfaces),
a ferromagnetic metal layer assembles both magnetism and spin-orbit
coupling, hence offering an alternative switching mechanism:\cite{manchon-prb,others}
Spin-orbit coupling transfers the orbital angular momentum carried by an electric current
to the electronic spin, thus creating an effective magnetic field (Rashba field).
As long as the effective field is mis-aligned with the magnetization direction,
the so-called Rashba torque emerges, thus exciting the magnetization.

Current-driven magnetization dynamics by spin-orbit torque has
been demonstrated by several experiments
on metal-oxide based systems. \cite{mihai1,pi,suzuki}
In fact, the Rashba torque can be categorized into to a broader
family of spin-orbit interaction induced torque that
has been observed in diluted magnetic semiconductors. \cite{chernyshov,fang,endo}
Recently, Miron \etal \cite{mihai3} has demonstrated the current-induced
magnetization switching using a \ti{single} ferromagnet in Pt/Co/AlO$_x$ trilayers, which further
consolidates the feasibility of the Rashba torque.
The same type of spin-orbit coupling induced torque is
predicted to improve current-driven domain wall
motion,\cite{tatara,others} which is supported by experimental observations. \cite{mihai2}
At this stage, we are aware of an alternative explanation,
as pointed out by Liu \etal \cite{liu} in terms of the 
spin Hall effect (SHE) occurring in the underlying heavy
metal layer, such as Pt or Ta. The distinction between the spin Hall induced effect
and the Rashba one is discussed in the last
section of this article.

In searching for a general form of the Rashba torque in
ferromagnetic metal layers,\cite{manchon-prb}
we found an expression that consists of
two components: \cite{wang-manchon-2011}An 
in-plane torque ($\propto{\bm m}\times({\hat{\bm y}}\times{\bm m})$) 
and an out-of-plane one ($\propto{\hat{\bm y}}\times{\bm m}$),
given ${\hat{\bm y}}$ is the in-plane direction transverse 
to the injected current and ${\bm m}$ is the magnetization direction. 
Numerical solution on a two-dimensional nano-wire with one open transport
direction has been carried out to appreciate the significance of diffusive
motion on the spin torque. We found that the in-plane component of the torque
increases when narrowing the magnetic wire\cite{wang-manchon-2011}.\par
 
In the present article, we give a full theoretical derivation of the coupled
diffusive equation for spin dynamics in a ferromagnetic metal layer and describe the form
of the Rashba torque in both weak and strong Rashba limits. In Sec. \ref{sec:diffusion-equation},
we combine the Keldysh formalism and the gradient expansion technique to
derive a coupled diffusion equation for charge and non-equilibrium spin densities.
To demonstrate that the diffusion equation
provides a coherent framework to describe the spin
dynamics, we dedicate Sec. \ref{sec:diff-ferro} to the spin diffusion in a ferromagnetic metal,
which shows an excellent agreement to early result on the same system.
In Sec.\ref{sec:she}, we illustrate that the absence of magnetism (in our diffusion equation)
describes the well-know phenomenon of electrically induced spin polarization.
The cases of a weak and a strong spin-orbit coupling are discussed in Sec.\ref{sec:weak-soi} and
Sec. \ref{sec:strong-soi}, respectively, where we provide an analytical form of the Rashba torque
in an infinite medium. In Sec. \ref{sec:discussion}, we discuss the implication
of the Rashba torque on magnetization dynamics as well as its distinction from
spin Hall effect induced torque.

\section{Diffusion equations}
\label{sec:diffusion-equation}
The system of interest is defined as a quasi-two-dimensional ferromagnetic metal layer
rolled out in the $xy$-plane. Two asymmetric interfaces
provide a confinement in $z$-direction, along which the potential gradient
generates a Rashba spin-orbit coupling. Therefore a single-particle
Hamiltonian for an electron of momentum $\hat{\bm{k}}$ is
($\hbar=1$ is assumed throughout)
\begin{align}
{\hat H}=\frac{\hat{\bm{k}}^{2}}{2m}+\alp \hat{\bm{\sg}}\cdot(\hat{\bm{k}}\times\hat{\bm{z}})
+\frac{1}{2}\Dlt_{xc} \hat{\bm{\sg}}\cdot\bm{m}+H^{i}
\end{align}
where $\hat{\bm{\sg}}$ is the Pauli matrix, $m$ the effective mass, and
$\bm{m}$ the magnetization direction. The ferromagnetic exchange splitting is
given by $\Dlt_{xc}$ and $\alp$ represents the Rashba constant (parameter).
Hamiltonian ${\hat H}^{i}=\sum_{j=1}^{N}V(\bm{r}-\bm{r}_{j})$ sums the contribution of the
non-magnetic impurity scattering potential $V(\bm{r})$ localized at $\bm{r}_{j}$.

To derive a diffusion equation for the non-equilibrium
charge and spin densities, we employ Keldysh
formalism. \cite{rammer-smith-rmp-1986}
Using Dyson equation, in a 2$\times$2 spin space, we obtain a kinetic equation that assembles the
retarded (advanced) Green's function $\hat{G}^{R}$ ($\hat{G}^{A}$),
the Keldysh component of the Green function
$\hat{G}^{K}$, and the self-energy $\hat{\Sg}^{K}$, \ie
\begin{align}
[\hat{G}^{R}]^{-1}\hat{G}^{K}-\hat{G}^{K}[\hat{G}^{A}]^{-1}
=\hat{\Sg}^{K}\hat{G}^{A}-\hat{G}^{R}\hat{\Sg}^{K},
\label{eq:kinetic-equation}
\end{align}
where all Green's functions are full functions with
interactions taken care of by the self-energies $\hat{\Sg}^{R,A,K}$.
The retarded (advanced)
Green's function in momentum and energy space is
\begin{align}
\hat{G}^{R(A)}(\bm{k},\eps)=\frac{1}
{\eps-\eps_{\bm{k}}-\hat{\bm{\sg}}\cdot\bm{b}(\bm{k})-\hat{\Sg}^{R(A)}(\bm{k},\eps)},
\end{align}
where $\eps_{\bm{k}}=\bm{k}^{2}/(2m)$ is the single-particle energy.
We have introduced a $\bm{k}$-dependent effective field
$\bm{b}(\bm{k})=\Dlt_{xc}\bm{m}/2+\alp(\bm{k}\times\bm{z})$ of the
magnitude $b_{k}=|\Dlt_{xc}\bm{m}/2+\alp(\bm{k}\times\bm{z})|$ and
the direction $\hat{\bm{b}}=\bm{b}(\bm{k})/b_{k}$.

Neglecting localization effect and electron-electron interactions,
we assume a short-range $\dlt$-function type
impurity scattering potential. At a low concentration and a weak coupling to electrons,
the second-order Born approximation is justified,\cite{rammer-smith-rmp-1986}
\ie the self-energy is \cite{mishchenko-prl-2004}
\begin{align}
\hat{\Sg}^{R,A,K}(\bm{r},\bm{r}^{\pr})
=\frac{\dlt(\bm{r},\bm{r}^{\pr})}{m\tau}\hat{G}^{R,A,K}(\bm{r},\bm{r})
\end{align}
where the momentum relaxation time reads
\begin{align}
\frac{1}{\tau}\approx 2\pi\int \frac{d^{2}\bm{k}^{\pr}}{(2\pi)^{2}}
|V(\bm{k}-\bm{k}^{\pr})|^{2} \dlt(\eps_{\bm{k}}-\eps_{\bm{k}^{\pr}}),
\end{align}
where $V(\bm{k})$ is the Fourier transform of the scattering potential and
the magnitude of momentum $\bm{k}$ and $\bm{k}^{\pr}$ is evaluated at
Fermi vector $k_{F}$.

The quasi-classical distribution function
$\hat{g}\equiv\hat{g}_{\bm{k},\eps}(T,\bm{R})$,
defined as the Wigner transform of the Keldysh function
$\hat{G}^{K}(\bm{r},t;\bm{r}^{\pr},t^{\pr})$, is obtained
by integrating out the relative spatial-temporal coordinates while retaining the
center-of-mass ones $\bm{R}=(\bm{r}+\bm{r}^{\pr})/2$ and
$T=(t+t^{\pr})/2$.  As long as the spatial profile of the quasi-classical
distribution function is smooth at the scale of Fermi wave length,
we may apply the gradient expansion technique
on Eq.(\ref{eq:kinetic-equation}),\cite{rammer-book}
which gives us a transport equation associated with macroscopic quantities.
The left-hand side of the kinetic equation in gradient expansion becomes
\begin{align}
[\hat{G}^{R}]^{-1}\hat{G}^{K}-\hat{G}^{K}[\hat{G}^{A}]^{-1} & \nn\\
\approx [\hat{g},\hat{\bm{\sg}}\cdot\bm{b}(\bm{k})]
& +\frac{i}{\tau}\hat{g}
+i\frac{\pat\hat{g}}{\pat T}\nn\\
+\frac{i}{2} & \left\{\frac{\bm{k}}{m}
+\alp(\hat{\bm{z}}\times\hat{\bm{\sg}}),\nabla_{\bm{R}}\hat{g}\right\},
\end{align}
where $\{\cdot,\cdot\}$ denotes the anti-commutator.
The relaxation time approximation indulges the
right-hand side of Eq.(\ref{eq:kinetic-equation}) as
\begin{align}
& \hat{\Sg}^{K}\hat{G}^{A}  -\hat{G}^{R}\hat{\Sg}^{K}\nn\\
& \approx  \frac{1}{\tau}\left[\hat{\rho}(\eps,T,\bm{R})\hat{G}^{A}(\bm{k},\eps)
-\hat{G}^{R}(\bm{k},\eps)\hat{\rho}(\eps,T,\bm{R})\right]
\end{align}
where we have introduced the density matrix by
integrating out the momentum $\bm{k}$ in $\hat{g}$, \ie
\begin{align}
\hat{\rho}(E,T,\bm{R})
=\frac{1}{2\pi N_{0}}\int \frac{d^{2}\bm{k}^{\pr}}{(2\pi)^{2}}
\hat{g}_{\bm{k}^{\pr},\eps}(T,\bm{R}).
\end{align}
For the convenience of discussion,
time variable is changed from $T$ to $t$.
At this stage, we have a kinetic equation depending on
$\hat{\rho}$ as well as on $\hat{g}$
\begin{align}
& i[\hat{\bm{\sg}}\cdot\bm{b}(\bm{k}),\hat{g}]
+\frac{1}{\tau}\hat{g}
+\frac{\pat\hat{g}}{\pat
t}+\frac{1}{2}\left\{\frac{\bm{k}}{m}
+\alp(\hat{\bm{z}}\times\hat{\bm{\sg}}),\nabla_{\bm{R}}\hat{g}\right\}\nn\\
& =\frac{i}{\tau}\left[\hat{G}^{R}(\bm{k},\eps)\hat{\rho}(\eps)
-\hat{\rho}(\eps)\hat{G}^{A}(\bm{k},\eps)\right].
\end{align}
A Fourier transformation
on temporal variable to the frequency domain $\og$ leads to
\begin{align}
\Og\hat{g}-b_{k}[\hat{U}_{k},\hat{g}]
=i\hat{K},
\label{eq:qke-with-K}
\end{align}
where $\Og=\og+i/\tau$ and the operator $\hat{U}_{k}\equiv \hat{\bm{\sg}}\cdot\hat{\bm{b}}$ satisfies
$\hat{U}_{k}\hat{U}_{k}=1$. The right hand side of Eq.(\ref{eq:qke-with-K}) is partitioned
according to
\begin{align}
\hat{K} =& \ub{-\frac{1}{2}\left\{\frac{\bm{k}}{m}
+\alp(\hat{\bm{z}}\times\hat{\bm{\sg}}),\nabla_{\bm{R}}\hat{g}\right\}}_{\hat{K}^{(1)}}\nn\\
&+\ub{\frac{i}{\tau}\left[\hat{G}^{R}(\bm{k}, \eps)\hat{\rho}(\eps)
-\hat{\rho}(\eps)\hat{G}^{A}(\bm{k},\eps)\right]}_{\hat{K}^{(0)}}.
\end{align}
The equilibrium part is denoted by $\hat{K}^{(0)}$ while the gradient term $\hat{K}^{(1)}$
is regarded as perturbation.
Functions $\hat{g}$ and $\hat{\rho}$ are both in frequency domain.
We solve Eq. (\ref{eq:qke-with-K}) formally to find a solution to $\hat{g}$
\begin{align}
\hat{g} = i\frac{(2 b_{k}^{2}-\Og^{2})\hat{K}
+2 b_{k}^{2}\hat{U}_{k}\hat{K}\hat{U}_{k}-\Og b_{k}[\hat{U}_{k},\hat{K}]}
{\Og(4 b_{k}^{2}-\Og^{2})}\equiv \mathscr{L}[\hat{K}].
\label{eq:equation-g}
\end{align}
An iteration procedure to solve Eq.(\ref{eq:equation-g})
has been outlined by Mishchenko \etal in Ref.[\onlinecite{mishchenko-prl-2004}].
We follow this procedure here: According to the partition scheme on $\hat{K}$,
we use $\hat{K}^{(0)}$ to obtain the zero-th order
approximation as $\hat{g}^{(0)}\equiv \msc{L}[\hat{K}^{(0)}(\hat{\rho})]$, which
replaces $\hat{g}$ in $\hat{K}^{(1)}$ to generate a correction due to
the gradient term, \ie $\hat{K}^{(1)}(\hat{g}^{(0)})$. We further
insert $\hat{K}^{(1)}(\hat{g}^{(0)})$ back to
Eq.(\ref{eq:equation-g}) to obtain a correction given by $\msc{L}[\hat{K}^{(1)}(\hat{g}^{(0)})]$,
then we obtain the first order approximation to the quasi-classical distribution function,
\begin{align}
\hat{g}^{(1)}=\hat{g}^{(0)}+\msc{L}[\hat{K}^{(1)}(\hat{g}^{(0)})].
\end{align}
The above procedure is repeated to any desired order, \ie
\begin{align}
\hat{g}^{(n)}=\hat{g}^{(n-1)}+\msc{L}[\hat{K}^{(1)}(\hat{g}^{(n-1)})].
\end{align}
In this paper, the second order approximation is sufficient.
The full expression of the second order approximation is tedious thus not included in the following.
A diffusion equation is derived by an angle averaging in momentum space,
which allows all terms that are odd order in $k_{i}$ ($i=x,y$) to vanish while the combinations such as
$k_{i}k_{j}$ contribute to the averaging by a factor $k_{F}^{2}\dlt_{ij}$,
given $k_{F}$ the Fermi wave vector.\cite{rammer-book}
Further more, a Fourier transform from frequency domain back to the real time
brings a diffusion type equation for the density matrix,
\begin{widetext}
\begin{align}
\frac{\pat}{\pat t}\hat{\rho}(t) = &D\nabla^{2}\hat{\rho}
-\frac{1}{\tau_{xc}}\hat{\rho}+\frac{1}{2\tau_{xc}}
(\hat{\bm{z}}\times\hat{\bm{\sg}})\cdot\hat{\rho}(\hat{\bm{z}}\times\hat{\bm{\sg}})
+iC \left[\hat{\bm{z}}\times\hat{\bm{\sg}},\bm{\nabla}\hat{\rho}\right]
-B \left\{\hat{\bm{z}}\times\hat{\bm{\sg}},\bm{\nabla}\hat{\rho}\right\}\nn\\
&+\Gamma \left[(\bm{m}\times\bm{\nabla})_{z}\hat{\rho}
-\hat{\bm{\sg}}\cdot\bm{m}\bm{\nabla}\hat{\rho}\cdot(\hat{\bm{z}}\times\hat{\bm{\sg}})
-(\hat{\bm{z}}\times\hat{\bm{\sg}})\cdot\bm{\nabla}\hat{\rho}\hat{\bm{\sg}}\cdot\bm{m}\right]\nn\\
&+\frac{1}{2 T_{xc}}\left(\hat{\bm{\sg}}\cdot\bm{m}\hat{\rho}\hat{\bm{\sg}}\cdot\bm{m}-\hat{\rho}\right)
-i\tilde{\Dlt}_{xc}[\hat{\bm{\sg}}\cdot\bm{m},\hat{\rho}]
-2R \left\{\hat{\bm{\sg}}\cdot\bm{m},(\bm{m}\times\bm{\nabla})_{z}\hat{\rho}\right\},
\label{eq:eom-leading-order}
\end{align}
\end{widetext}
where all quantities are evaluated at Fermi energy $\eps_{F}$.
In a two-dimensional system, the diffusion constant $D=\tau v_{F}^{2}/2$ is given
in terms of Fermi velocity $v_{F}$ and momentum relaxation time $\tau$.
The renormalized exchange splitting reads $\tilde{\Delta}_{xc}=(\Dlt_{xc}/2)/(4\xi^2+1)$ where
$\xi^{2}=(\Dlt_{xc}^{2}/4+\alp^{2}k_{F}^{2})\tau^{2}$.
The other parameters are
\begin{align}
& C =\frac{\alp k_{F}v_{F}\tau}{(4\xi^{2}+1)^{2}},\;
\Gamma=\frac{\alpha\Dlt_{xc}v_Fk_F\tau^2}{2(4\xi^2+1)^2},\;
R =\frac{\alp\Dlt_{xc}^{2}\tau^{2}}{2(4\xi^{2}+1)},\nn\\
& ~\frac{1}{\tau_{xc}}=\frac{2\alp^{2}k_{F}^{2}\tau}{4\xi^{2}+1},\;
B =\frac{2\alpha^3 k_F^2\tau^2}{4\xi^2+1},
\frac{1}{T_{xc}}=\frac{\Dlt_{xc}^2\tau}{4\xi^2+1}.\nn
\end{align}
$\tau_{xc}$ is the relaxation time due to the so-called D'yakonov-Perel mechanism.\cite{dp-she-1971}
Equation (\ref{eq:eom-leading-order}) is valid in the dirty
limit $\xi\ll 1$, which enables the approximation
$1+4\xi^{2}\approx 1$.
Charge density $n$ and the non-equilibrium spin density $\bm{S}$ are introduced
by the vector decomposition on the density matrix $\hat{\rho}=n/2+\bm{S}\cdot{\hat{\bm\sg}}$.
In a real experimental setup, \cite{mihai1,mihai2,mihai3} spin transport in ferromagnetic layers
suffers from random magnetic scatterers, for which we introduce
an isotropic spin-flip relaxation $\bm{S}/\tau_{sf}$ phenomenologically.

Eventually, we obtain a set of diffusion equations for the charge and
spin densities, \ie
\begin{align}
\frac{\pat n}{\pat t} =& D\nabla^{2}n+ B {\bm\nabla}_z\cdot{\bm S}\nn\\
& +\Gamma {\bm\nabla}_z\cdot {\bm m} n
+R  {\bm\nabla}_z\cdot \bm{m}(\bm{S}\cdot \bm{m}),
\label{eq:charge-diffusion}
\end{align}
and
\begin{align}
\frac{\pat \bm {S}}{\pat t} =& D \nabla^{2}{\bm S}
-\frac{1}{\tau_{\para}}\bm{S}_{\para}
-\frac{1}{\tau_{\perp}}\bm{S}_{\perp}\nn\\
&-\Dlt_{xc}\bm{S}\times \bm{m}
-\frac{1}{T_{xc}} \bm{m}\times(\bm{S}\times \bm{m})\nn\\
&+B {\bm\nabla}_z n
+2 C {\bm\nabla}_z\times{\bm S}
+2 R ( \bm{m}\cdot{\bm\nabla}_z n) \bm{ m}\nn\\
&+\Gamma \left[ \bm{m}\times({\bm\nabla}_z\times{\bm S})
+{\bm\nabla}_z\times({\bm m}\times{\bm S})\right],
\label{eq:spin-dynamics}
\end{align}
where $\bm{\nabla}_{z}\equiv \hat{\bm{z}}\times\bm{\nabla}$.
The spin density $\bm{S}_{\para}\equiv S_{x}\hat{\bm{x}}+S_{y}\hat{\bm{y}}$
is relaxed at a rate  $1/\tau_{\para}\equiv 1/\tau_{xc}+1/\tau_{sf}$ while
$\bm{S}_{\perp}\equiv S_{z}\hat{\bm{z}}$ has a rate $1/\tau_{\perp}\equiv 2/\tau_{xc}+1/\tau_{sf}$.

For a broad range of the relative strength between spin-orbit coupling and the
exchange splitting, \ie $\alp k_{F}/\Delta_{xc}$,
Eq.(\ref{eq:charge-diffusion}) and Eq.(\ref{eq:spin-dynamics})
describe the spin dynamics in a ferromagnetic layer.
When the magnetism vanishes ($\Dlt_{xc}=0$), the $B$-term provides a source that generates
spin density electrically. \cite{edelstein-1989,mishchenko-prl-2004}
On the other hand, when the spin-orbit coupling is absent ($\alpha=0$),
the first two lines in Eq.(\ref{eq:spin-dynamics})
describe a diffusive motion of spin density in a ferromagnetic metal,
which, to be shown in the next section, agrees excellently with early results
in the corresponding limit. \cite{zhang-li-2004}
$C$-term describes the coherent precession of the
spin density around the effective Rashba field.
The precession of the spin density (induced by the Rashba field) around the exchange
field is described by the $\Gamma$-term, thus a higher order (compared to
$C$) in the dirty limit for $\Gamma=\Dlt_{xc}\tau C/2$. The $R$-term contributes to the
magnetization renormalization.

\section{Spin diffusion in a ferromagnet}
\label{sec:diff-ferro}
Spin diffusion in a ferromagnet has been discussed actively
in the field of spintronics. \cite{zhang-levy-fert-2002,zhang-li-2004,tserk-rmp-2005,tserk-prb-2009}
In this section we show explicitly that, by suppressing the spin-orbit coupling,
Eq.(\ref{eq:spin-dynamics}) describes the spin diffusion equation in the corresponding limits.

In the present model, vanishing Rashba spin-orbit coupling means $\alp=0$, then
Eq.(\ref{eq:spin-dynamics}) reduces to
\begin{align}
\frac{\pat}{\pat t}\bm{S} = D\nabla^{2}\bm{S}+ & \frac{1}{\tau_{\Dlt}}\bm{m}\times\bm{S}\nn\\
-&\frac{1}{\tau_{sf}}\bm{S}
- \frac{1}{T_{xc}}{\bm m}\times({\bm S}\times{\bm m}),
\end{align}
where $\tau_{\Dlt}\equiv 1/\Dlt_{xc}$ is the time scale of the coherent precession of the
spin density around the magnetization.
This equation differs from the result of Zhang \etal \cite{zhang-levy-fert-2002}
only by a dephasing of the transverse component of the spin density
that is set by the time scale $T_{xc}$.
In a ferromagnetic metal, we may divide the spin density into
a \ti{longitudinal} component that follows the magnetization direction adiabatically,
and a deviation that is \ti{perpendicular} to the magnetization, \ie
$\bm{S}=s_{0}\bm{m}+\dlt\bm{S}$ where $s_{0}$ is the local equilibrium spin
density. Such a partition, after restoring the electric field by
$\bm{\nabla}\raw\bm{\nabla}+e\bm{E}\pat_{\eps}$, gives rise to
\begin{align}
\frac{\pat}{\pat t}\dlt\bm{S}+\frac{\pat}{\pat t}s_{0}\bm{m}& \nn\\
=s_{0}D\nabla^{2}\bm{m}&+D\nabla^{2}\dlt\bm{S}+D e P_{F}\mac{N}_{F}\bm{E}\cdot\bm{\nabla}\bm{m}\nn\\
-\frac{\dlt\bm{S}}{\tau_{sf}}&-\frac{s_{0}\bm{m}}{\tau_{sf}}-\frac{\dlt\bm{S}}{T_{xc}}
+\Dlt_{xc}\bm{m}\times\dlt\bm{S},
\label{eq:full-spinaccu-dynamics}
\end{align}
where the magnetic order parameter is allowed to be
spatial dependent, \ie $\bm{m}=\bm{m}({\bm r},t)$.
The energy derivative is treated as
$\pat_{\eps}\bm{S}\approx P_{F}\mac{N}_{F}\bm{m}$ given $P_{F}$ the spin
polarization and $\mac{N}_{F}$ the density of state, both at Fermi energy.

For a smooth magnetic texture in which the characteristic length scale of the
magnetic profile is much larger than the length scale for electron transport,
we discard the contribution $D\nabla^{2}\dlt\bm{S}$. \cite{zhang-li-2004}
The diffusion of the equilibrium spin density follows
$s_{0}D\nabla^{2}\bm{m}\approx s_{0}\bm{m}/\tau_{sf}$. In this paper, we
retain only terms that are first order in temporal
derivative, which simplifies Eq.(\ref{eq:full-spinaccu-dynamics}) to
\begin{align}
\label{eq:zl1}
-\frac{1}{\tau_{\Dlt}}\bm{m}\times\dlt\bm{S}
+&\left(\frac{1}{\tau_{sf}}+\frac{1}{T_{xc}}\right)\dlt\bm{S}= \nn\\
-&s_{0}\frac{\pat}{\pat t}\bm{m}
+D e P_{F}\mac{N}_{F}\bm{E}\cdot\bm{\nabla}\bm{m}.
\end{align}
The last equation can be solved exactly
\begin{align}
\label{eq:zl2-spin-accu}
\dlt\bm{S}=&\frac{\tau_{\Dlt}}{1+\varsigma^2}
\left[\frac{P_F}{e}{\bm m}\times({\bm j}_e\cdot{\bm\nabla}){\bm m}
+\varsigma \frac{P_F}{e}({\bm j}_e\cdot{\bm\nabla}){\bm m}\right.\nn\\
&\left.-s_{0}{\bm m}\times\frac{\pat\bm{m}}{\pat t}-\varsigma s_{0}\frac{\pat\bm{m}}{\pat t}\right]
\end{align}
where $\varsigma=\tau_\Dlt(1/\tau_{sf}+1/T_{xc})$ and the electric
current $\bm{j}_{e}=e^{2}n\tau\bm{E}/m$ is given in terms of electron density
$n$. Apart from the inclusion of the dephasing of
transverse component as implemented in parameter $\varsigma$,
the non-equilibrium spin density Eq.(\ref{eq:zl2-spin-accu})
agrees excellently with Eq.(8) in Ref.[\onlinecite{zhang-li-2004}].

Given the knowledge of the spin density, the spin torque, defined as
\begin{align}
{\bm T}=-\frac{1}{\tau_{\Dlt}}\bm{m}\times\dlt{\bm S}+\frac{1}{T_{xc}}\dlt{\bm S},
\label{eq:spin-torque-definition}
\end{align}
is given by
\begin{align}
\label{eq:zl2-torque}
{\bm T}=&\frac{1}{1+\varsigma^2}
\left[-\eta s_{0}\frac{\pat\bm{m}}{\pat t}
+\beta s_{0}{\bm m}\times\frac{\pat\bm{m}}{\pat t}\right.\nn\\
&\left.+\eta \frac{P_F}{e}({\bm j}_e\cdot{\bm\nabla})\bm{m}
-\beta \frac{P_F}{e}\bm{m}\times({\bm j}_e\cdot{\bm\nabla})\bm{m}\right]
\end{align}
where $\eta=1+\varsigma\tau_{\Dlt}/T_{xc}$ and $\beta=\tau_{\Dlt}/\tau_{sf}$.
Assuming a long dephasing time of the transverse component
(\ie $T_{xc}\raw\infty$), then $\eta\approx 1$
and Eq. (\ref{eq:zl2-torque}) reproduces the Eq.(9) in
Ref.[\onlinecite{zhang-li-2004}]. On the other hand,
a short dephasing time (of the transverse component)
enhances parameter $\eta$ therefore increases the temporal
spin torque (\ie the first term in Eq.(\ref{eq:zl2-torque})).

\section{Electrically generated spin density}
\label{sec:she}
The effect of an electrically generated non-equilibrium
spin density due to spin-orbit coupling \cite{edelstein-1989}
can be extracted from Eq.(\ref{eq:spin-dynamics}) by
setting exchange interaction zero (\ie $\Dlt_{xc}=0$). Retaining D'yakonov-Perel as the only
spin relaxation mechanism and letting $\tau_{sf}=\infty$, Eq.(\ref{eq:spin-dynamics}) ends up in
\begin{align}
D\bm{\nabla}^2\bm{S}-& \frac{1}{\tau_{xc}}({\bm S}+S_z{\hat{\bm z}})\nn\\
&+2C ({\hat{\bm z}}\times{\bm\nabla})\times{\bm S}
+B ({\hat{\bm z}}\times{\bm\nabla})n=0
\label{eq:spin-hall-effect}
\end{align}
which reduces to the results in the well-known spin Hall effect.
\cite{mishchenko-prl-2004,burkov-prb-2004,adagideli-bauer-prl-2005}
In the case of an infinite medium along transport direction, \ie $\hat{\bm{x}}$-direction,
Eq.(\ref{eq:spin-hall-effect}) gives rise to a solution to the spin density
\begin{align}
{\bm S}=&\tau_{xc}B e E \frac{1}{\eps_{F}} n\hat{\bm{y}}
=\frac{e E\zeta}{\pi v_{F}}\hat{\bm{y}},\nn
\end{align}
where only the linear term in electric field has been retained. On the right hand side,
we have used the charge density in a 2D system $n=k_{F}^{2}/(2\pi)$ and
introduced the parameter $\zeta=\alp k_{F}\tau$
as used in Ref. [\onlinecite{mishchenko-prl-2004}].

In the following sections, we explore the spin
torque in the presence of both exchange and Rashba
field in an infinite medium. The primary focus is on two cases:
Weak and a strong spin-orbit coupling,
when comparing to the magnitude of exchange splitting.
In general, Eq.(\ref{eq:spin-dynamics}) is applicable through a broad range
of relative strength between spin-orbit coupling and exchange splitting.
A full scale numerical simulation on the diffusion
equation is beyond the scope of this paper,
we refer the readers to Ref.[\onlinecite{wang-manchon-2011}] for further interests.

\section{Weak spin-orbit coupling}
\label{sec:weak-soi}
A weak Rashba spin-orbit coupling implies a small D'yakonov-Perel relaxation
rate $1/\tau_{xc}\propto \alp^{2}$, such that $\tau_{xc}\gg \tau_{sf}, \tau_{\Dlt}$,
which allows spin relaxation to be dominated by random magnetic impurities.
In this regime, when comparing to the magnitudes of $C$ and $\Gamma$,
the contribution from $B$ and $R$ are at a higher order in $\alp$, thus to be disregarded.
We consider a stationary state where $\pat\bm{S}/\pat t=0$.
An electric field applied along ${\hat{\bm x}}$-direction, \ie $\bm{E}=E\hat{\bm{x}}$.
In an infinite medium,\cite{manchon-prb} all the spatial derivatives
vanishes ($\bm{\nabla}\raw 0$) and the dynamic equation reads
\begin{align}
-\frac{1}{\tau_\Dlt}&{\bm m}\times{\bm S}
+\frac{1}{T_{xc}}{\bm m}\times({\bm S}\times{\bm m})
+\frac{1}{\tau_{sf}}{\bm S}\nn\\
=& 2 e E C{\hat{\bm y}}\times\partial_\eps{\bm S}\nn\\
&+eE\Gamma\left[{\hat{\bm y}}\times({\bm m}\times\partial_\eps{\bm S})
+{\bm m}\times({\hat{\bm y}}\times\partial_\eps{\bm S})\right].
\label{eq:weak-alp-general}
\end{align}
In addition to the spin density induced by exchange splitting,
a weak spin-orbit interaction leads to a deviation in spin density that can be
considered as a perturbation. Therefore, we may well apply the partition
${\bm S}={\bm S}_\perp+S_{\para}\bf{m}$ to separate the longitudinal and
the transverse components. Eq.(\ref{eq:weak-alp-general}) is thus reduced to
\begin{align}
\frac{1}{\tau_\Dlt}{\bm m}\times{\bm S}_\bot
-\frac{1}{T_{\perp}}{\bm S}_{\perp}-\frac{1}{\tau_{sf}}S_{\para}\bm{m}&\nn\\
=-2eECP_{F}\mac{N}_{F}& \hat{\bm{y}}\times\bm{m}\nn\\
-eE\Gamma P_{F}& \mac{N}_{F}\bm{m}\times(\hat{\bm{y}}\times\bm{m})
\label{eq:perp-spin-accu-weak-alp}
\end{align}
where $1/T_{\perp}\equiv 1/T_{xc}+1/\tau_{sf}$ and we have again employed
the approximation on the energy derivative
$\pat_{\eps}\bm{S}\approx P_{F}\mac{N}_{F}\bm{m}$ and replaced the
energy derivative of the charge density by the density of states at Fermi energy
(\ie $\pat_{\eps} n\approx n/\eps_{F}=\mac{N_{F}}$).
We solve Eq.(\ref{eq:perp-spin-accu-weak-alp}) to obtain a solution to the
non-equilibrium spin density
\begin{align}
\bm{S}_{\perp}& =\frac{\tau_\Dlt}{1+\varsigma^2}e E P_{F}\mac{N}_{F}
\left[(2C+\varsigma\Gamma){\bm m}\times(\hat{\bm y}\times{\bm m})\right.\nn\\
&\left.-(\Gamma-2\varsigma C)(\hat{\bm y}\times{\bm m})\right].
\label{eq:spin-density-weak-soi}
\end{align}
and $\bm{S}_{\para}=0$. In Eq.(\ref{eq:spin-density-weak-soi}),
the second component, oriented along the direction $\hat{\bm{y}}\times\bm{m}$,
is actually perpendicular to the plane spanned by the magnetization direction
and the effective Rashba field (along $\hat{\bm{y}}$), which, as to be shown below,
contributes to a Rashba torque that fulfils
the symmetry described in a recent experiment.\cite{mihai3}
The definition Eq.(\ref{eq:spin-torque-definition})
leads to a general expression for the spin torque
\begin{align}
\bm{T}=&T_{\perp}\hat{\bm{y}}\times\bm{m}+T_{\para}\bm{m}\times(\hat{\bm{y}}\times\bm{m}),
\label{eq:torque-weak-soi}
\end{align}
which consists of an \ti{out-of-plane} and an \ti{in-plane} components with
magnitudes determined by
\begin{align}
T_\perp=&\frac{e E P_{F}\mac{N}_F}{1+\varsigma^2}(2 \eta C+\beta\Gamma),\\
T_\para=&\frac{e E P_{F}\mac{N}_F}{1+\varsigma^2}(\eta\Gamma-2 \beta C).
\label{eq:torque-magnitude-weak-alp}
\end{align}
The \ti{plane} is defined by the magnetization direction $\bm{m}$ and the
direction of the effective Rashba field that in the present setting is aligned
along $\hat{\bm{y}}$-direction. Note that the sign of the in-plane torque, 
Eq. (\ref{eq:torque-magnitude-weak-alp}), can change depending on the 
interplay between spin relaxation and precession.

To compare directly with the results in Ref.[\onlinecite{manchon-prb}],
we allow the spin relaxation time $\tau_{sf}\raw \infty$,
therefore $\beta\approx 0$. We also consider the transverse dephasing time
to be infinite.\cite{zhang-levy-fert-2002,zhang-li-2004} Under these assumptions,
$\eta\approx 1$ and $\varsigma\approx 0$ and we have
$T_\perp\approx 2 e E P_{F}\mac{N}_F C$ and
$T_\para\approx e E P_{F}\mac{N}_F\Gamma$. In the dirty limit,
$\Gamma\ll C$ due to $\Dlt_{xc}\tau\ll 1$, therefore
making use of the relation for the polarization
$P_{F} = \Dlt_{xc}/\eps_{F}$ and the Drude
relation $j_{e}=e^{2}n\tau E/m$, we obtain an out-of-plane torque
\begin{eqnarray}
\bm{T}
=2\frac{\alp m\Dlt_{xc}}{e\eps_{F}}j_{e}\hat{\bm{y}}\times\bm{m},
\end{eqnarray}
which agrees excellently with the spin torque in an infinite system in the
corresponding limit as derived in Ref.[\onlinecite{manchon-prb}].

\section{Strong spin-orbit coupling}
\label{sec:strong-soi}
The opposite limit to Sec.\ref{sec:weak-soi} is a strong spin-orbit coupling.
In this case, we consider the scenario that $\alp k_{F}\gg \Dlt_{xc}$ and the D'yakonov-Perel
relaxation mechanism is dominating, \ie $1/\tau_{xc}\gg 1/\tau_{sf}$, due to
the fact $1/\tau_{xc}\propto \alp^{2}$.
Therefore, it is not physical to simply assume that the direction of spin
density is dominantly aligned along the magnetization direction, as what is treated
in the case of a weak spin-orbit coupling.
A self-consistent solution from Eq.(\ref{eq:spin-dynamics}) to the spin density is more justified.

Again, as in Sec.\ref{sec:weak-soi}, we consider an infinite system where an electric
field $\bm{E}$ is applied at $\hat{\bm{x}}$-direction. The magnetization direction is left arbitrary.
We approximate the energy derivative by $\pat_{\eps}\approx 1/\eps_{F}$.
The above assumptions simplify Eq.(\ref{eq:spin-dynamics}) to
\begin{align}
\frac{1}{\tau_\Dlt}\bm{S}\times\bm{m}
+ \frac{1}{T_{xc}}&\bm{m}\times({\bm S}\times \bm{m})
-\frac{2e E C}{\eps_{F}}\hat{\bm{y}}\times\bm{S}\nn\\
+ & \frac{1}{\tau_{xc}}({\bm S}+S_{z}\hat{\bm{z}})
= \frac{e E}{\eps_{F}} n B \hat{\bm{y}},
\label{eq:largerashba}
\end{align}
where a strong spin-orbit coupling renders $\Gamma$ and $R$ terms negligible.
By considering $T_{xc}\gg\tau_{\Dlt},\tau_{xc}$, Eq. (\ref{eq:largerashba}) reduces to
\begin{align}
\frac{1}{\tau_\Dlt}\bm{S}\times{\hat {\bm m}}
+\frac{1}{\tau_{xc}}({\bm S}+S_{z}\hat{\bm{z}})&\nn\\
-\frac{2e E C}{\eps_{F}}\hat{\bm{y}}\times\bm{S}
& =\frac{e E}{\eps_{F}}n B \hat{\bm{y}},
\end{align}
which is a set of linear equations for the non-equilibrium spin density.
We are interested in the linear response regime, which implies that
at the distance as defined by the Fermi wave length
$1/k_{F}$, we have $eE/k_{F}\ll \alp k_{F}$.
Therefore up to the first order in exchange splitting, we extract the spin density
from the above equation to be
\begin{align}
{\bm S}=\frac{eE}{\eps_F}n \tau_{xc}B \left({\hat{\bm y}}
-\chi \hat{\bm{y}}\times\bm{m}-\frac{\chi}{2}m_x{\hat {\bm z}}\right)
\end{align}
where $\chi\equiv \tau_{xc}/\tau_{\Dlt}$
we have used the identity $\hat{\bm{y}}\times\bm{m}=m_{z}\hat{\bm{x}}-m_{x}\hat{\bm{z}}$.
This yields a spin torque
\begin{align}
{\bm T}=\frac{\alp m \Dlt_{xc}}{e\eps_{F}}j_{e} & \left({\hat {\bm y}}\times{\bm m}\right.\nn\\
&\left.+\chi{\bm m} \times({\hat{\bm y}}\times{\bm m})
-\frac{\chi}{2}m_x{\hat{\bm z}}\times {\bm m}\right).
\label{eq:torque-strong-rashba}
\end{align}
This torque is slightly different from the weak
Rashba limit and has a strong implication in
terms of magnetization dynamics. The torque is
dominated by a field-like torque along $\hat{\bm{y}}$,
similarly to the weak Rashba case. First, in contrast to the 
weak Rashba case [see Eq. (\ref{eq:torque-magnitude-weak-alp})], 
the sign of the in-plane torque remains positive. Secondly,
the anisotropic spin relaxation coming from
D'yakonov-Perel mechanism yields an additional component of spin accumulation
that is oriented along ${\hat{\bm z}}$. The implication of this torque on the
current-driven magnetization dynamics is discussed in the next section.

\section{Discussion}
\label{sec:discussion}
Current-induced magnetization dynamics in a single ferromagnetic layer
has been observed in various structures that involve interfaces between
transition metal ferromagnets, heavy metals and/or metal-oxide insulators.
Existing experimental systems are 
Pt/Co/AlO$_x$, \cite{mihai1,mihai2,mihai3,pi} Ta/CoFeB/MgO, \cite{suzuki}
Pt/NiFe and Pt/Co bilayers. \cite{liu} 
Besides the structural complexity in such systems, 
an unclear form of spin-orbit coupling in the bulk and interfaces places a challenge to  
understand the nature of the torque.

\subsection{Validity of Rashba model}
The celebrated Rashba-type effective interfacial spin-orbit Hamiltonian was pioneered by
E. I. Rashba to model the influence of
asymmetric interfaces in semiconducting 2DEG: \cite{rashba-soi}
A sharp potential drop, emerging at the interface (say, in the $xy$-plane)
between two materials, gives rise to a potential gradient ${\bm\nabla}V$ that is
normal to the interface, \ie $\bm{\nabla}V\approx \xi(\bm{r})\hat{\bm{z}}$.
In case a rotational symmetry exists in the interface plane,  
a spherical Fermi surface assumption allows the spin-orbit interaction
Hamiltonian to have the form
${\hat H}_{R}=\alp \hat{\bm{\sg}}\cdot(\bm{p}\times{\hat{\bm z}})$,
where $\alp \approx\langle\xi\rangle/4m^2c^2$. 
As a matter of fact, in semiconducting interfaces where the transport 
is described by a limited number of bands around a high symmetry point, 
the Rashba form can be recovered through $\bm{k}\cdot\bm{p}$ theory. \cite{winkler}

As far as metallic interfaces are concerned, a spin-orbit splitting of the 
Rashba-type in the conduction band has been observed
at Au surfaces, \cite{metals} Gd/GdO interfaces, \cite{Gd} Bi surfaces and
compounds, \cite{Bi} and metallic quantum wells. \cite{qw} The presence of a Rashba
interaction in graphene \cite{graphene} and at oxide hetero-interfaces \cite{oxides}
has also been  reported recently. It is quite interesting to notice that the symmetry breaking-induced
spin splitting of the conduction band seems rather general and might not be
restricted to heavy metal interfaces \cite{qw}.

In the case of transition metals, however, the free electron approximation fails
to characterize the band structure accurately due to both a large number of band
crossing at the Fermi energy and a strong hybridization among $s$, $p$ and $d$ orbitals.
Density functional theory (DFT) is a successful tool to investigate the nature of
spin-orbit interaction at metallic surfaces. For example, in Refs.[\onlinecite{bilh}],
the authors observe a band splitting that possesses similar properties as Rashba 
spin-orbit interaction and decays exponentially away from the surface. \cite{bilh} 
Alternatively, the spin-orbit interaction at metallic surfaces has been addressed using
tight-binding models for the $p$ orbitals. \cite{petersen,park}
At such sharp interfaces, the magnitude of the orbital angular momentum (OAM) 
is considered to play a dominant role at the onset of a Rashba-type spin splitting.

This finding is consistent with the long standing work on interfacial
magnetic anisotropy at a ferromagnet/heavy metal,\cite{ani} and more recently,
ferromagnetic/metal-oxide interface.\cite{hongxi} In such systems,
a perpendicular magnetic anisotropy arises from the orbital overlap between
the 3$d$ states of the ferromagnets and the spin-orbit coupled states of the normal metal.
The observation of perpendicular magnetic anisotropy at Co/metal-oxide interfaces
tends to support the major role of large interfacial OAM in the onset of interfacial
spin-orbit effects. \cite{park,hongxi} The presence of interfacial Rashba spin-orbit
coupling has also been shown to produce interfacial
perpendicular magnetic anisotropy. \cite{manchon}

All these previous theoretical and experimental studies strongly suggest that the interfacial spin splitting exists
in the presence of a large OAM and potential
gradient. However, a microscopic description of realistic interfaces
is still missing. Although the Rashba spin-orbit interaction is a
convenient Hamiltonian to extract qualitative behaviors, its
applicability to realistic metallic interfaces with complex band structures
remains to be tested.

\subsection{Spin Hall effect versus Rashba torque}
Recently, Liu \etal \cite{liu} proposed to manipulate the magnetization
of a Pt/Co or Pt/NiFe bilayer using the spin current generated by
spin Hall effect in the underlying Pt layer. When injecting a charge
current $\bm{j}_e$ into a normal metal accommodating a strong spin-orbit coupling,
the asymmetric spin scattering induces a transverse pure spin current that has the
form ${\cal J}=(\alpha_H/e)\bm{j}_e\times {\hat{\bm \mu}}\otimes{\hat{\bm \mu}}$,
where $\alpha_H$ is the spin Hall angle and ${\hat{\bm \mu}}$ is the
spin direction.\cite{she} When impinging on the ferromagnetic layer deposited
on top of the Pt layer, the spin current transverse to the local magnetization
is absorbed and generates a torque 
$\bm{T}_{SHE}=(b_H/e)(1-\beta\bm{m}\times)\bm{m}\times({\hat{\bm \mu}}\times{\bm m})$
(to be called SHE torque thereafter).
Here, $b_H=\alpha_Hj_e\mu_B/e$ is the spin torque amplitude 
where the regular spin polarization $P$ is replaced by the spin 
Hall angle $\alpha_H$. $\beta$ is the non-adiabaticity parameter
proposed by Zhang and Li \cite{zhang-li-2004}
and it stems from the presence of spin-flip scattering in the system.
In the configuration adopted by Liu \etal
the charge current is injected along $\hat{\bm{x}}$ and the torque is 
given by 
\begin{align}
{\bm T}_{SHE}=\alpha_H\mu_B\frac{j_e}{e}\left({\bm m}\times({\hat{\bm y}}\times{\bm m})
+\beta{\hat{\bm y}}\times{\bm m}\right).
\end{align}
Note that a more realistic model should account for spin diffusion in Co and Pt,
as discussed in Ref. [\onlinecite{vedy}]. An important conclusion is that, besides
the correction in the case of a strong Rashba coupling, both Rashba and SHE
produce the same type of torque, see Eq.(\ref{eq:torque-weak-soi}) and 
Eq.(\ref{eq:torque-strong-rashba}) in this article. 

Nevertheless, distinctions can be made. First, in the absence
of the corrections due to spin-flip and spin precession, the Rashba torque reduces to
the field-like term, ${\hat{\bm y}}\times{\bm m}$, whereas the SHE torque
reduces to the (anti-)damping term ${\bm m}\times({\hat{\bm y}}\times{\bm m})$.
This assertion must be scrutinizede carefully since the actual relative magnitude
between the field-like and the damping torques depends on the width of the magnetic wire
as well as on the detailed spin dynamics in presence of spin-flip and precession. \cite{wang-manchon-2011}
Furthermore, for such an ultra-small system the spin-flip scattering giving rise
to the non-adiabaticity parameter ($\beta$) might be significantly different from the one
measured in a more conventional thin film.

A second important difference arises from the fact that the Rashba torque arises from
spin-orbit fields generated by {\em interfacial} currents, whereas the SHE
torque is due to the current flowing in the {\em bulk} of the Pt layer.
Therefore, for a constant external electric field, varying the thickness of
Pt layer shall enhance the SHE torque, while keeping the Rashba torque unchanged.

The torques as a function of the Co layer thickness is more difficult to foresee.
Although one could claim that Rashba spin-orbit interaction is expected to be
localized at the interface, where the potential gradient is large, numerical simulations
show that the Rashba-type interaction survives a few monolayers \cite{bilh}
(which is typically the thickness of the Co layer under consideration).
In addition, the presence of quantum well states might also modify the nature
of the spin-orbit interaction in the ultrathin magnetic layer in a
system such as Pt/Co/AlO$_x$. \cite{qw}

The same is true for the SHE torque. The injection of spin current
into a Co layer is accompanied by spin precession that takes place over a
very short decoherence length. This decoherence length has been studied
experimentally and theoretically in spin valves and found to be of the
order of a few monolayers. \cite{decoh} In the typical case of 3 or 4
monolayer-thick ferromagnets, the SHE torque can not be considered as a
purely interfacial phenomenon.

\subsection{Magnetization Dynamics}
In Pt/Co/AlO$_x$ trilayers, Miron \ti{et al} have observed a current-driven domain
wall nucleation, \cite{mihai1} an enhanced current-driven domain wall
velocity \cite{mihai2} and a current-driven magnetization switching.\cite{mihai3}
The symmetry of the spin torque required to explain the
experimental findings agree well with Rashba torque proposed in Ref. \onlinecite{manchon-prb}.
On a similar structure, Pi \ti{et al} \cite{pi} and Suzuki \ti{et al} \cite{suzuki} also observed
an effective field torque that could be interpreted in terms of the Rashba torque.
Recently, Liu \ti{et al} \cite{liu} interpreted their experiments on
Pt/NiFe and Pt/Co bilayers using SHE in the underlying Pt layer.

\subsubsection{Magnetization switching}
According to our previous discussions, both Rashba torque and
SHE torque have a general form
${\bm T}=T_\perp{\hat{\bm y}}\times{\bm m}+T_\para{\bm m}\times({\hat{\bm y}}\times{\bm m})$.
The first term acts like a field oriented along the direction transverse
current direction whereas the second term acts like an (anti-)damping term,
mimicking a conventional spin transfer torque that would arise from a
polarizer pointing to ${\hat{\bm y}}$.

As a consequence, both Rashba torque and SHE torque possess the
appropriate symmetry to excite the magnetization of a single
ferromagnet and induce switching, as observed by 
Miron \ti{et al} \cite{mihai3} and Liu \ti{et al}. \cite{liu}
In the case of a large Rashba spin-orbit coupling, the torque acquires
an additional component that acts like an effective magnetic
field along ${\hat{\bm z}}$, vanishing as the
magnetization component $m_x$ is zero (see Section \ref{sec:strong-soi}),
which provides an additional torque that helps destabilize the magnetization.

\subsubsection{Current-driven domain wall motion}
The influence of Rashba/SHE torque on a domain wall can be
illustrated within the rigid Bloch wall approximation.
The perpendicularly magnetized Bloch wall is parameterized
by ${\bm m}=(\cos\phi\sin\theta,\sin\phi\sin\theta,\cos\theta)$
where $\phi=\phi(t)$ and $\theta(x,t)=2\tan^{-1}[e^{(x-x_c(t))/\Delta}]$,
where $x_c$ refers to the center of the domain wall and $\Delta$ is
defined as the domain wall width. To describe the dynamics of a Bloch wall, 
Landau-Lifshitz-Gilbert (LLG) equation 
\begin{align}
\partial_t{\bm m}=-\gamma{\bm m}\times{\bm H}_{eff}
+\alp_{G}\partial_t{\bm m}\times{\bm m}+{\bm \tau}
\label{eq:llg1}
\end{align}
has to be augmented by the current induced torque $\bm{\tau}$ 
\begin{align}
\bm{\tau}= b_J{\bm \nabla}{\bm m}-\beta b_J{\bm m}\times{\bm \nabla}{\bm m}&\nn\\
+b_J(\tau_\perp{\hat{\bm y}}\times{\bm m}+\tau_\para{\bm m} & \times({\hat{\bm y}}\times{\bm m})\nn\\
&+\tau_z m_x{\hat{\bm z}}\times{\bm m}).
\label{eq:llg3}
\end{align}
The torque $\bm{\tau}$ is written in the most general form,
where the first two terms are the regular adiabatic and the 
so-called non-adiabatic torques; the next two terms ($\tau_\para$ and $\tau_\perp$)
emerge from the presence of Rashba and/or spin Hall effect
and the last term $\tau_z$ appears only in large Rashba 
limit (see Sec. \ref{sec:strong-soi}). The magnitude of 
the adiabatic torque is $b_J=\mu_BPj_e/e$. The effective field is given by
\begin{align}
\bm{H}_{eff}=\frac{2A}{M_s}{\bm\nabla}^2{\bm m}
+H_{K}m_x{\hat{\bm x}}+H_{\perp}m_z{\hat{\bm z}}.
\label{eq:llg2}
\end{align}
Parameter $\gamma$ in LLG is the gyromagnetic ratio, $\alp_{G}$ is the
Gilbert damping, $A$ is the exchange constant, $M_s$ is the
saturation magnetization, $H_K$ is the in-plane magnetic anisotropy
and $H_\perp$ is the combination of an out-of-plane anisotropy and
a demagnetizing field. The magnetization dynamics can be obtained readily from Eqs. (\ref{eq:llg1})-(\ref{eq:llg2}) 
by integrating over the magnetic volume
\begin{align}
\label{eq:llg4}
\pat_t\phi+\alpha_{G}\frac{\pat_t x_c}{\Dlt} 
=&\left[\frac{\Delta\pi}{2}(\tau_{\para}-\frac{\tau_z}{2})\cos\phi-\beta\right]\frac{b_J}{\Delta}\\
\alpha_{G}\partial_t\phi-\frac{\partial_t x_c}{\Delta}=& -\gm \frac{H_K}{2}
 \sin2\phi+\left(1+\frac{\Delta\pi}{2}\tau_{\perp}\cos\phi\right)\frac{b_J}\Delta.
\label{eq:llg5}
\end{align}
We observe that the in-plane torque $\tau_\para$
distorts the domain wall texture, while the perpendicular torque
$\tau_\perp$ drives the domain wall motion. The additional torque
$\tau_z$, arising in the large Rashba limit, only contributes to the in-plane torque.
Therefore, in the following, we will refer to the in-plane torque
as $\tau_\para^{\ast}=\tau_{\para}-\tau_z/2$. Below the Walker
breakdown ($\partial_t\phi=0$), the velocity is given by
\begin{align}\label{eq:il}
&\pat_t x_c=-\left(\beta-\frac{\Dlt \pi}{2}\tau_\para^{\ast}\cos\phi\right)\frac{b_J}{\alpha_{G}}\\
&\gm\frac{H_K}{2}\sin2\phi
=\left[\alpha_{G}-\beta+\frac{\Delta\pi}{2}(\alpha_{G}\tau_\perp
+\tau_\para^{\ast})\cos\phi\right]\frac{b_J}{\alpha_G\Delta},
\label{eq:il2}
\end{align}
where the tilting angle $\phi$ is given by the competition between
the magnetic anisotropy, the non-adiabatic torque, and the Rashba/SHE torque.
In the case of weak Rashba ($\tau_z=0$), assuming $\tau_\para=\beta\tau_\perp$
and omitting the correction to the spin precession, we recover the results of Ref. [\onlinecite{kim}].
When neglecting the in-plane torque and accounting for the
perpendicular Rashba torque ($\tau_\para^{\ast}=0$), the Rashba torque only acts like an
effective transverse field and enhances the Walker breakdown limit \cite{mihai2} [see Eq. (\ref{eq:il2})].

Accounting for the in-plane component $\tau_\para$ arising either
from corrections to Rashba torque or from the SHE,
this torque appears to modify the domain wall velocity.
Therefore, depending on the strength and the sign of Rashba/SHE
torque as well as on the resulting tilting angle $\phi$, it is possible
to obtain a vanishing or even a reversed domain wall velocity, as has been 
shown numerically in Ref. [\onlinecite{kim}] and illustrated in Eq. (\ref{eq:il}).
A full scale numerical investigation is beyond the scope of this article,
but it will help understand the profound effect of Rashba and
SHE torque on the domain wall structures.

\section{Conclusion}
Using Keldysh technique, in the presence of both magnetism and a Rashba spin-orbit
coupling, we derive a spin diffusion equation that provides a coherent
description to the diffusive spin dynamics. 
In particular, we have derived a general expression for the 
Rashba torque in the bulk of a ferromagnetic metal layer, at both weak and strong Rashba limits.
We find that the torque is in general composed of two components, a 
field-like torque and the other (anti-)damping one.
Being aware of the recent alternative interpretation on the current-induced magnetization switching
in a single ferromagnet, we have discussed the difference between the Rashba and the 
SHE torques. While exploring the common features, 
we found that the magnetization dynamics driven by the Rashba torque 
presents several interesting similarities to that induced by SHE torque. 
Nevertheless, further investigation involving
structural modification of the system is expected to provide a deeper knowledge 
on the nature of the interfacial spin-orbit interaction as well as the current-induced 
magnetization switching in a single ferromagnet.

\begin{acknowledgments}
We thank G. E. W. Bauer,
J. Sinova, M. D. Stiles, X. Waintal, and S. Zhang
for stimulating discussions. We are specially grateful to K. -J. Lee and H. -W. Lee for inspiring discussion about the magnetization dynamics.
\end{acknowledgments}


\begin{thebibliography} {999}
\bibitem{dp-she-1971}
M. I. D'yakonov and V. I. Perel, Sov. Phys. JETP Lett.
{\bf 13}, 467 (1971); Phys. Lett. A {\bf 35}, 459 (1971).
\bibitem{edelstein-1989}
V. M. Edelstein, Solid. Stat. Comm. {\bf 73}, 233 (1989).
\bibitem{murakami-science-2004}
S. Murakami \etal Science {\bf 301}, 1348 (2004).
\bibitem{sinova-prl-2004}
J. Sinova \ti{et al.}, Phys. Rev. Lett. {\bf 92}, 126603 (2004).
\bibitem{kato-science-2004}
Y. K. Kato \etal Science {\bf 306}, 1910 (2004).
\bibitem{rashba-soi}
Yu. A. Bychkov and E. I. Rashba, J. Phys. C: Solid. Stat. Phys. {\bf 17}, 6039 (1984).
\bibitem{dp}
M. I. D'yakonov and V. I. Perel, Sov. Phys. Solid State {\bf 13}, 3023 (1971).
\bibitem{slonczewski-berger-1996}
J. C. Slonczewski, J. Magn. Magn. Mater. {\bf 159}, L1 (1996);
L. Berger, Phys. Rev. B. {\bf 54}, 9353 (1996).
\bibitem{refstt}
M. D. Stiles and J. Miltat, Top. Appl. Phys. {\bf 101}, 225 (2006);
D. C. Ralph and M. D. Stiles, J. Magn. Magn. Mater. {\bf 320}, 1190 (2008);
J. Z. Sun and D. C. Ralph, \ti{ibid}. {\bf 320}, 1227 (2008).
\bibitem{manchon-prb}
A. Manchon and S. Zhang, Phys. Rev. B. {\bf 78}, 212405 (2008);
\ti{ibid}. {\bf 79}, 094422 (2009).
\bibitem{others}
A. Matos-Abiague and R. L. Rodriguez-Suarez, Phys. Rev. B {\bf 80}, 094424 (2009);
I. Garate and A. H. MacDonald, \ti{ibid}. {\bf 80}, 134403 (2009);
P. M. Haney, and M. D. Stiles, Phys. Rev. Lett. {\bf 105}, 126602 (2010).
\bibitem{mihai1}
I. M. Miron \etal  Nature Materials {\bf 9}, 230 (2010).
\bibitem{pi}
U. H. Pi \etal Appl. Phys. Lett. {\bf 97}, 162507 (2010).
\bibitem{suzuki}
T. Suzuki \etal Appl. Phys. Lett. {\bf 98}, 142505 (2011).
\bibitem{mihai3}
I. M. Miron \etal  Nature (London) {\bf 476}, 189 (2011).
\bibitem{chernyshov}
A. Chernyshov \etal Nature Physics {\bf 5}, 656 (2010).
\bibitem{fang}
D. Fang \etal  Nature Nanotechnology {\bf 6}, 413 (2011).
\bibitem{endo}
M. Endo \etal Appl. Phys. Lett. {\bf 97}, 222501 (2010).
\bibitem{tatara}
K. Obata, and G. Tatara, Phys Rev. B {\bf 77}, 214429 (2008).
\bibitem{mihai2}
I. M. Miron \etal  Nature Materials {\bf 10}, 419 (2011).
\bibitem{liu}
L. Liu \etal Phys. Rev. Lett. {\bf 106}, 036601 (2011); arXiv:1110.6846 (2011).
\bibitem{wang-manchon-2011}
X. Wang and A. Manchon, arXiv:1111.1216 (2011).
\bibitem{rammer-smith-rmp-1986}
J. Rammer and H. Smith, Rev. Mod. Phys. {\bf 58}, 323 (1986).
\bibitem{mishchenko-prl-2004}
E. G. Mishchenko \etal Phys. Rev. Lett. {\bf 93}, 226602 (2004).
\bibitem{rammer-book}
J. Rammer, \ti{Quantum Field Theory of Non-equilibrium States}
(Cambridge University Press, Cambridge, 2007).
\bibitem{zhang-li-2004}
S. Zhang and Z. Li, Phys. Rev. Lett. {\bf 93}, 127204 (2004).
\bibitem{zhang-levy-fert-2002}
S. Zhang \etal Phys. Rev. Lett. {\bf 88}, 236601 (2002).
\bibitem{tserk-rmp-2005}
Y. Tserkovnyak \etal Rev. Mod. Phys. {\bf 77}, 1375 (2005).
\bibitem{tserk-prb-2009}
Y. Tserkovnyak \etal Phys. Rev. B {\bf 79}, 094415 (2009).
\bibitem{burkov-prb-2004}
A. A. Burkov \etal Phys. Rev. B {\bf 70}, 155308 (2004).
\bibitem{adagideli-bauer-prl-2005}
\.I. Adagideli and G. E. W. Bauer, Phys. Rev. Lett. {\bf 95}, 256602 (2005).
\bibitem{winkler}
R. Winkler, {\em Spin-Orbit Coupling Effects in Two-Dimensional
Electron and Hole Systems}, Springer, Berlin, (2003).
\bibitem{metals}
S. LaShell \etal Phys. Rev. Lett. {\bf 77}, 3419 (1996);
F. Reinert \etal Phys. Rev. B {\bf 63}, 115415 (2001);
G. Nicolay \etal \ti{ibid}. {\bf 65}, 033407 (2001);
M. Hoesch \etal \ti{ibid}. {\bf 69}, 241401(R) (2004).
\bibitem{Gd}
O. Krupin \etal Phys. Rev. B {\bf 71}, 201403(R) (2005);
O Krupin \etal  New. J. Phys. {\bf 11}, 013035 (2009).
\bibitem{Bi}
Ph. Hofmann, Prog. Surf. Sci. {\bf 81}, 191 (2006);
T. Hirahara \etal Phys. Rev. B {\bf 76}, 153305 (2007);
H. Mirhosseini \etal \ti{ibid}. {\bf 79}, 245428 (2009);
A. Takayama \etal Phys. Rev. Lett. {\bf 106}, 166401 (2011);
K. Ishizaka \etal Nature Materials {\bf 10}, 521 (2011).
\bibitem{qw}
A. Varykhalov \etal Phys. Rev. Lett. {\bf 101}, 256601 (2008);
J. Hugo Dil \etal \ti{ibid}. {\bf 101}, 266802 (2008);
A. G. Rybkin \etal Phys. Rev. B {\bf 82}, 233403 (2010).
\bibitem{graphene}
Yu. S. Dedkov \etal Phys. Rev. Lett. {\bf 100}, 107602 (2008);
O. Rader \etal \ti{ibid}. {\bf102}, 057602 (2009).
\bibitem{oxides}
A. D. Caviglia \etal Phys. Rev. Lett. {\bf 104}, 126803 (2010).
\bibitem{bilh}
G. Bihlmayer \etal Phys. Rev. B {\bf 75}, 195414 (2007);
G. Bihlmayer \etal Surf. Sci. {\bf 600}, 3888 (2006);
M. Nagano \etal J. Phys.: Cond. Mater. {\bf 21}, 064239 (2009).
\bibitem{petersen}
L. Petersen and P. Hedeg\aa rd, Surf. Sci. {\bf 459}, 49 (2000).
\bibitem{park}
S. R. Park \etal Phys. Rev. Lett. {\bf107}, 156803 (2011).
\bibitem{ani}
K. Kyuno \etal J. Phys. Soc. Jpn. {\bf 61}, 2099 (1992);
D. Weller \etal Phys. Rev. B {\bf 49}, 12888 (1994);
G. H. O. Daalderop \etal \ti{ibid}. {\bf 50}, 9989 (1994).
\bibitem{hongxi}
H. X. Yang \etal Phys. Rev. B {\bf 84}, 054401 (2011);
C. Nistor \etal \ti{ibid}. {\bf 84}, 054464 (2011).
\bibitem{manchon}
A. Manchon, Phys. Rev. B {\bf 83}, 172403 (2011).
\bibitem{she}
J. E. Hirsch, Phys. Rev. Lett. {\bf 83}, 1834 (1999); 
S. Zhang, \ti{ibid}. {\bf 85}, 393 (2000).
\bibitem{vedy}
A. Vedyayev \etal arXiv:1108.2589 (2011).
\bibitem{decoh}
S. Wang \etal Phys. Rev. B {\bf 77}, 184430 (2008).
\bibitem{kim}
K.-W. Kim \etal arXiv:1111.3422 (2011).
\end{thebibliography}
\end{document}